\documentclass[11pt,a4paper]{article}
\pdfoutput = 1
\usepackage{jheppub}
\usepackage{color}
\usepackage{amsmath,bm}
\usepackage{graphicx}

\newcommand{\as}{\alpha_{\mathrm{s}}}
\newcommand{\LA}{\mathrm{A}}
\newcommand{\LB}{\mathrm{B}}

\newcommand{\LI}{\mathrm{I}}
\newcommand{\LJ}{\mathrm{J}}

\newcommand{\LR}{\mathrm{R}}

\newcommand{\LT}{\mathrm{T}}
\newcommand{\LZ}{\mathrm{Z}}
\newcommand{\La}{\mathrm{a}}
\newcommand{\Lb}{\mathrm{b}}

\newcommand{\Lf}{\mathrm{f}}
\newcommand{\Lg}{\mathrm{g}}

\newcommand{\z}{z}

\newcommand{\GeV}{\ \mathrm{GeV}}

\definecolor{red}{rgb}{1,0,0}

\def\<>#1{\big\langle{#1}\big\rangle}
\def\[]#1{\big[{#1}\big]}

\newbox\charbox
\newbox\slabox
\def\s#1{{      
        \setbox\charbox=\hbox{$#1$}
        \setbox\slabox=\hbox{$/$}
        \dimen\charbox=\ht\slabox
        \advance\dimen\charbox by -\dp\slabox
        \advance\dimen\charbox by -\ht\charbox
        \advance\dimen\charbox by \dp\charbox
        \divide\dimen\charbox by 2
        \raise-\dimen\charbox\hbox to \wd\charbox{\hss/\hss}
        \llap{$#1$}
}}

\title{Ordering variable for parton showers}

\author[a]{Zolt\'an Nagy}

\author[b]{and Davison E.\ Soper}

\affiliation[a]{
DESY\\
Notkestrasse 85\\
22607 Hamburg, Germany
}

\affiliation[b]{
Institute of Theoretical Science\\
University of Oregon\\
Eugene, OR  97403-5203, USA
}

\emailAdd{Zoltan.Nagy@desy.de}
\emailAdd{soper@uoregon.edu}

\abstract{
The parton splittings in a parton shower are ordered according to an ordering variable, for example the transverse momentum of the daughter partons relative to the direction of the mother, the virtuality of the splitting, or the angle between the daughter partons. We analyze the choice of the ordering variable and conclude that one particular choice has the advantage of factoring softer splittings from harder splittings graph by graph in a physical gauge.
}

\keywords{perturbative QCD, parton shower}
\preprint{DESY 13-242}

\begin{document}
\maketitle

\section{Introduction}
\label{sec:intro}

In a companion paper \cite{deductor}, we have introduced a parton shower event generator, \textsc{Deductor} \cite{DeductorCode}, based on our earlier work \cite{NSI,NSII}. The corresponding shower algorithm is designed to be suitable for an improved treatment of spin, as described in ref.~\cite{NSspin}, and for an improved treatment of color, as described in ref.~\cite{NScolor}. This shower generator contains features that differ from other parton shower event generators even when one uses the leading color approximation and averages over spins, as we do in ref.~\cite{deductor}. One of these features is that the algorithm uses non-zero masses for initial state partons, which requires modified evolution equations for the parton distribution functions, as described in a separate companion paper \cite{MassivePdfs}. The second feature is the choice of shower evolution variable, which is the subject of this paper.

In a parton shower event generator, when a parton labelled $i$ splits to two partons, one typically assigns a variable $V_i^2$ to the splitting, where $V_i^2$ is a function of the momenta of the mother parton and its daughters. The purpose of defining $V_i^2$ is to order splittings within the shower: if the splitting of parton $i$ comes before the splitting of parton $j$ then $V_i^2 > V_j^2$. When $V_i^2$ has dimensions of mass squared, one can define a dimensionless shower time $t_i \propto \log(Q_0^2/V_i^2)$ where $Q_0^2$ denotes the scale of the hard interaction that initiates the shower. Then $t$ increases as the shower progresses. 

In most cases ($k_\LT$-ordering, virtuality ordering, but {\em not} angular ordering), $V_i^2$ is a measure of the hardness of the splitting: $V_i^2 \to 0$ when the angle between the daughter partons approaches zero or when the momentum of one of the daughter partons approaches zero. The ordering variable in \textsc{Deductor} is hardness based in this sense.

There is a physics reason for choosing a hardness based ordering parameter. The reason is simple to state if we are not to precise about it. We think of the simulated event as measured by possible physical observables.  We can imagine a class of observables that have a resolution scale corresponding to a certain value of the shower time. With observables in this class, we see what happens at space-time separations from the hard interaction that are smaller than some value $1/V_i$. Anything that happens at larger space-time separations is ``unresolvable.'' To measure an observable in this class, we need to run the parton shower to time $t_i \propto \log(Q_0^2/V_i^2)$, but no further. If we use an observable that has a resolution down to softer interactions, we see more detail, corresponding to greater space-time separations. The parton shower algorithm allows us to see more detail if we generate more splittings that are softer and softer.

\begin{figure}
\centerline{\includegraphics[width=9.0cm]{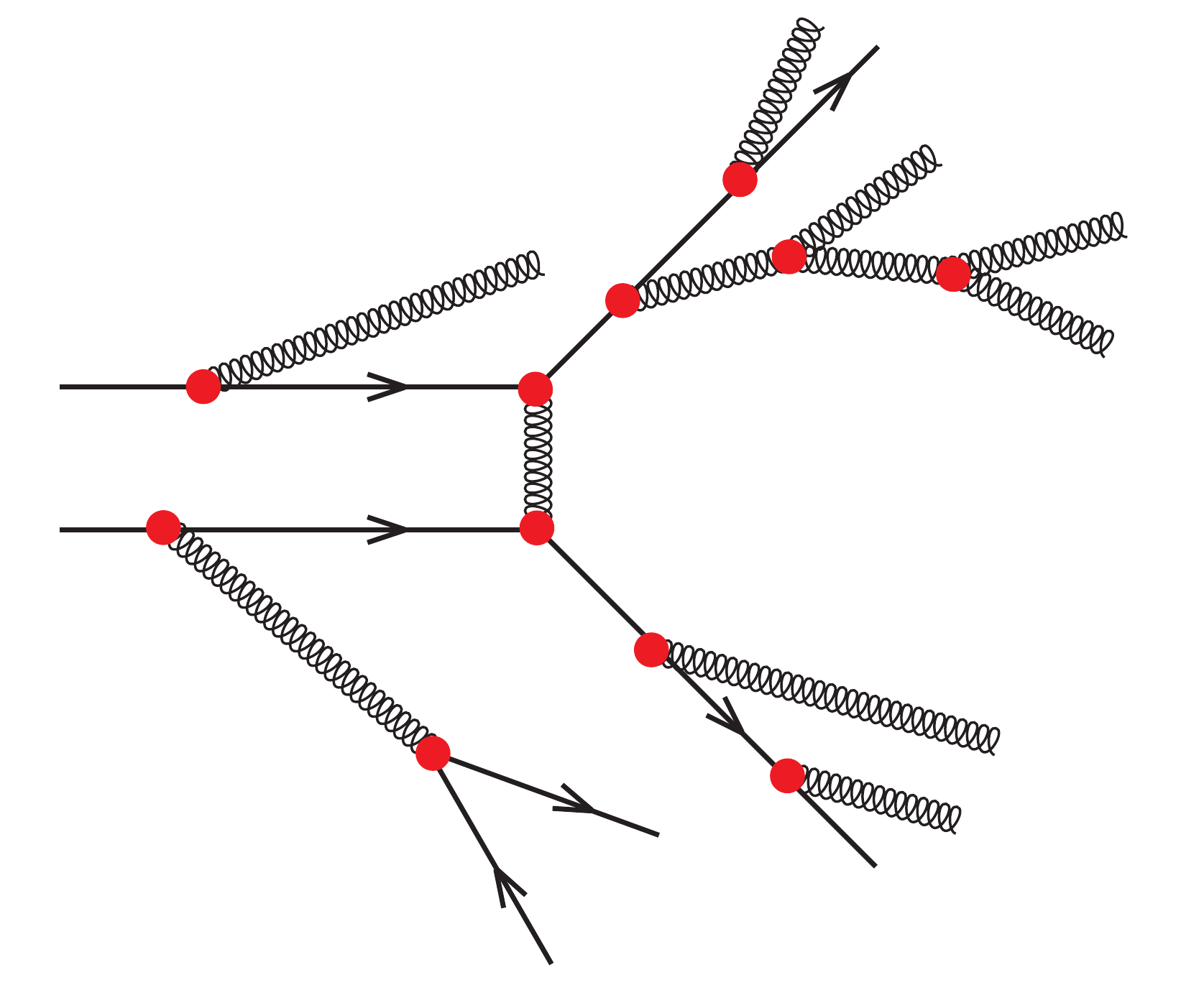}}
\caption{A possible shower history. The hard interaction is quark-quark scattering via gluon exchange. Both the initial state quarks and the final state quarks can emit gluons. Shower time increases as one moves away from the hard interaction. Parton splittings at later shower time are softer.}
\label{fig:showerhistories}
\end{figure}

What does this look like as realized in the parton shower event generator? The parton shower generates final states starting with some hard process in hadron-hadron scattering. The shower includes both final state splittings and initial state splittings from the incoming partons. For each event, there is a shower history, $H$, as in figure \ref{fig:showerhistories}. The event generator generates a given shower history with a probability $P_H$.

Now, the shower history has the appearance of a Feynman diagram, with its corresponding amplitude ${\cal M}_H$. Let's assume that we use a physical gauge for evaluating ${\cal M}_H$. One might imagine that $P_H$ is approximately proportional to the corresponding $|{\cal M}_H|^2$, $P_H \approx {\cal N}|{\cal M}_H|^2$. (The normalization factor ${\cal N}$ need not concern us here.) That can't be quite right because the event generator needs to account for interference graphs, in which a soft gluon is emitted from one parton in ${\cal M}_H$ and from a different parton in an amplitude ${\cal M}_{H'}^*$. For that reason, we really need to discuss amplitudes ${\cal M}_H$. We will, in fact, discuss amplitudes, but let us ignore this issue for a first orientation. Then it is possible, but not necessary, that the $P_H$ is an approximation to ${\cal N}|{\cal M}_H|^2$ for the corresponding Feynman diagram. It is not necessary that $P_H \approx {\cal N}|{\cal M}_H|^2$ because all that one really needs is that $\sum_H P_H$ approximates ${\cal N}\sum_H|{\cal M}_H|^2$. 

Nevertheless, we believe that it is desirable that $P_H$ approximates ${\cal N}|{\cal M}_H|^2$ (after accounting for interference graphs) and we used that idea as a design goal in constructing the parton shower algorithm used in \textsc{Deductor}. Maintaining this diagram by diagram correspondence constrains the choice of the ordering variable for shower evolution. Our purpose in this paper is to study this constraint and explain why we made the choice of shower time used in \textsc{Deductor}.

\section{Shower time for final state splittings}
\label{sec:finalstatesplittings}

In order to see why a correspondence between shower splitting probabilities and Feynman diagrams constrains the choice of shower time, it is easiest to start with final state splittings. Suppose that, in a shower history $H$ of interest, a final state parton 0 splits into partons 1 and 2, each of which splits further. This continues through some (finite) number of splittings. 

In the Feynman diagram corresponding to history $H$, the momenta of the mother parton and of the two daughters are related by $p_0 = p_1 + p_2$. Let the partons have masses $m_i$, $i = 0,1,2$. We define corresponding virtualities $v_i^2$ by $v_i^2 = p_i^2 - m_i^2$. With final state splittings, we have $v_i^2 \ge 0$. In our Feynman diagram, the propagator for the mother parton has a denominator $1/v_0^2$. The daughter partons have propagators with denominators $1/v_1^2$ and $1/v_2^2$.

With the idea that the shower evolves from hard interactions to softer interactions, 
the picture for the Feynman diagram is that the splitting $0 \to 1 + 2$ should be relatively hard compared to the subsequent splittings of the daughter partons 1 and 2. In particular, we ought to be able to set $v_1^2$ and $v_2^2$ to 0 when calculating the propagator denominator $1/[(p_1 + p_2)^2 - m_0^2] = 1/v_0^2$ that controls the probability of the splitting $0 \to 1 + 2$.\footnote{Large logarithms are a feature of parton showers. In this situation we generate a large logarithm from an integration $\int\!dv_0^2/v_0^2$, integrated over a wide range of $v_0^2$. A second power of the logarithm comes from integrating over the momentum fraction $\z$ in the splitting.} This statement is of some practical significance. When the shower algorithm generates the splitting $0 \to 1 + 2$, the splittings of partons 1 and 2 have not yet been generated. Thus $v_1^2$ and $v_2^2$ are unknown. Knowing only that the daughter virtualities should be small, we set them to zero.

This qualitative argument suggests a design feature for a parton shower. The definition of the shower ordering variable should be such that one can neglect $v_1^2$ and $v_2^2$ when calculating $v_0^2$.

We now need to make more precise the proposition that one should be able to ``neglect $v_1^2$ and $v_2^2$ when calculating $v_0^2$.'' We make use of the idea that parton showers are about the development of jets: groups of particles with large momenta approximately collinear to a certain direction. Letting $Q_0$ denote the total momentum of the final state partons created by the hard process that initiates the shower, we define components of vectors using a reference frame in which $Q_0$ has only a time component. We align the $z$-axis of our reference frame along the direction of the jet of interest.

We suppose that the mother parton is part of the jet of interest and we describe its momentum $p_0$ using $(+,-,\perp)$ components\footnote{Specifically, $v^\pm = (v^0 \pm v^3)/\sqrt 2$. Then $v^2 = 2 v^+ v^- - \bm v^2$.} in our reference frame. In this frame, partons in the jet have momenta with large plus components and small minus components: $p^+ \gg |\bm p| \gg p^-$, where the transverse components are denoted by a boldface symbol, $\bm p$. The components of $p_0$ are 
\begin{equation}
\label{eq:p0components}
p_0 = \left(P,\,\frac{\bm p_0^2 + m_0^2 + v_0^2}{2P},\,\bm p_0 \right)
\;\;.
\end{equation}
Here $m_0$ is the mass of the mother parton, $\bm p_0$ is its momentum transverse to the jet direction, and $v_0^2$ is its virtuality. With our choice of reference frame, $P$ is large: $P^2 \gg \bm p_0^2$, $P^2 \gg m_0^2$, $P^2 \gg v_0^2$. We could have chosen the reference frame so that $\bm p_0 = 0$, but we leave the choice general. Parton 0 splits into two partons with momenta
\begin{equation}
\begin{split}
\label{eq:p12components}
p_1 ={}& \left(\z P,\,\frac{\bm p_1^2 + m_1^2 + v_1^2}{2\z P},\,\bm p_1 \right)
\;\;,
\\
p_2 ={}& \left((1-\z) P,\,\frac{\bm p_2^2 + m_2^2 +v_2^2}{2(1-\z)P},\, \bm p_2 \right)
\;\;.
\end{split}
\end{equation}
Here $\z$ and $(1-\z)$ are the momentum fractions carried by the daughter partons. Their transverse momenta are related by
\begin{equation}
\bm p_1 + \bm p_2 = \bm p_0
\;\;.
\end{equation}
Partons 1 and 2 are off shell with virtualities $v_1^2$ and $v_2^2$, respectively. Setting $p_0^- = p_1^- + p_2^-$ and solving for $v_0$, we find
\begin{equation}
\label{eq:Q0sq}
v_0^2
=
\frac{((1-\z)\bm p_1 - \z \bm p_2)^2}
{\z(1-\z)}
+\frac{m_1^2}{\z} + \frac{m_2^2}{1-\z} - m_0^2
+\frac{v_1^2}{\z}
+\frac{v_2^2}{(1-\z)}
\;\;.
\end{equation}

If we neglect $v_1^2$ and $v_2^2$ when calculating $v_0^2$, we obtain $\tilde v_0^2$, where
\begin{equation}
\label{eq:Q0sqtilde}
\tilde v_0^2
=
\frac{((1-\z)\bm p_1 - \z \bm p_2)^2}
{\z(1-\z)}
+\frac{m_1^2}{\z} + \frac{m_2^2}{1-\z} - m_0^2
\;\;.
\end{equation}
The approximation $v_0^2 \approx \tilde v_0^2$ is a good approximation provided
\begin{equation}
\frac{v_1^2}{\z} \ll v_0^2\,,
\hskip 1 cm
\frac{v_2^2}{1-\z} \ll v_0^2
\,.
\end{equation}
That is 
\begin{equation}
\begin{split}
\label{eq:thecondition0}
\frac{v_1^2}{2 p_1\cdot n} \ll{}& \frac{v_0^2}{2 p_0\cdot n}
\;,
\hskip 1 cm
\frac{v_2^2}{2 p_2\cdot n} \ll \frac{v_0^2}{2 p_0\cdot n}
\;.
\end{split}
\end{equation}
where $n$ is the lightlike vector $n = (0,Q_0^-,\bm 0)$. This definition gives $n$ dimensions of momentum.

We want eq.~(\ref{eq:thecondition0}) to hold whenever the splittings of the two daughter partons come after the splitting $0 \to 1 + 2$. To guarantee that, we want to arrange the definition of shower time such that the splitting of parton 1 comes after the splitting of parton 0 when the first of conditions (\ref{eq:thecondition0}) holds and
such that the splitting of parton 2 comes after the splitting of parton 0 when second of conditions (\ref{eq:thecondition0}) holds. 

These ordering conditions are a simple restatement of the requirement that $v_0^2 \approx \tilde v_0^2$. But what is $\tilde v_0^2$? It is the limit of $v_0^2$ as $v_1^2 \to 0$ and $v_2^2 \to 0$ with fixed $\bm p_1$, $\bm p_2$, and $\z$. This was a deliberate choice. It was motivated as follows. We define $\tilde v_0^2 = (\tilde p_1 + \tilde p_2)^2 - m_0^2$, where $\tilde p_1$ and $\tilde p_2$ are on-shell approximations to $p_1$ and $p_2$: $\tilde p_i^2 = m_i^2$ for $i = 1,2$. We want the components of $\delta p_i = p_i - \tilde p_i$ to be small. Since $v_i^2 = (\tilde p_i^2 + \delta p_i)^2 - m_i^2 = 2 \tilde p_i \cdot \delta p_i + (\delta p_i)^2$, we have, neglecting $(\delta p_i)^2$,
\begin{equation}
v_i^2/2 \approx \tilde p_i^+ \delta p_i^- - \tilde {\bm p}_i \cdot \delta \bm p_i
+ \tilde p_i^- \delta p_i^+
\;.
\end{equation}
Now, $\tilde p_i^+ \gg |\tilde {\bm p}_i| \gg \tilde p_i^-$. Thus we can make the components of $\delta p_i$ small by choosing 
\begin{equation}
\label{eq:deltapminus}
\delta p^- = \frac{v_i^2}{2 \tilde p_i^+}
\end{equation}
and letting $\delta \bm p_i = 0$ and $\delta p_i^+ = 0$. One could also let $\delta \bm p_i$ and $\delta p_i^+$ be small, of a similar size to $\delta p^-$, but then $\delta p^-$ is still determined approximately by eq.~(\ref{eq:deltapminus}). Such a choice is equivalent to the simple choice that we make here.\footnote{For instance, when the partons are part of a high energy jet, if we define $\tilde v_0^2$ by taking the limit $v_1^2 \to 0$ and $v_2^2 \to 0$ with the three-momenta of the partons held constant, adjusting the daughter parton energies a little to put them on-shell, we obtain the same final definition of shower time.}

We need one more step in order to turn eq.~(\ref{eq:thecondition0}) into a definition of shower time. Since the plus components of the parton momenta $p_0, p_1$ and $p_2$ in this frame are all much larger than their transverse  and minus components, we have
\begin{equation}
p_i\cdot n \approx p_i\cdot Q_0
\;.
\end{equation}
Then we can write the ordering conditions (approximately) as
\begin{equation}
\begin{split}
\label{eq:thecondition1}
\frac{v_1^2}{2\,p_1\cdot Q_0} \ll {}& \frac{v_0^2}{2\,p_0\cdot Q_0}
\;,
\hskip 1 cm
\frac{v_2^2}{2\,p_2\cdot Q_0} \ll \frac{v_0^2}{2\,p_0\cdot Q_0}
\;.
\end{split}
\end{equation}

This leads us to the definition of shower time for a final state splitting. For the splitting of parton $i$, we take
\begin{equation}
\label{eq:tdef}
e^{-t_i} = \frac{\Lambda_i^2}{Q_0^2}
\end{equation}
where we define the ordering variable $\Lambda$ by
\begin{equation}
\label{eq:showertime}
\Lambda_i^2 = \frac{p_i^2 - m_i^2}{2\,p_i\cdot Q_0} \,Q_0^2
\;.
\end{equation}

With this definition, the splittings of partons 1 and 2 come after the splitting of parton 0 when the conditions (\ref{eq:thecondition1}) hold. Of course, within a parton shower we calculate $p_i^2$ as the square of the sum of the momenta of the daughter partons to parton $i$, using the approximation that the daughter partons are on shell. Additionally, at each splitting a small amount of momentum needs to be taken from the existing final state partons in the event, as described in section 4.1 of ref.~\cite{NSI}. Thus, there is an ambiguity in how to calculate $p_i\cdot Q_0$. We calculate $p_i\cdot Q_0$ defining $p_i$ to be the momentum of the parton $i$ as it existed when the parton was created, before it splits.

In the shower algorithm, we take ``after'' to mean $t_1 > t_0$ and $t_2 > t_0$. This replaces $\ll$ in eq.~(\ref{eq:thecondition1}) by simply $<$. This does not treat the regions $t_1 \approx t_0$ and $t_2 \approx t_0$ very exactly. We recognize this as a shortcoming of a leading order shower that can be fixed if we move on to a next-to-leading order shower.

\section{Shower time and physical time}

Using $(+,-,\perp)$ components as in eq.~(\ref{eq:p0components}), the shower time for the splitting of the mother parton 0 is
\begin{equation}
e^{-t} = \frac{v_0^2}{2 \,p_0\cdot Q_0}
\approx \sqrt{\frac{2}{Q_0^2}}\, \frac{v_0^2}{2P}
= \sqrt{\frac{2}{Q_0^2}}\,\left[
p_0^- - \frac{\bm p_0^2 + m_0^2}{2P}
\right]
\;.
\end{equation}
Here $({\bm p_0^2 + m_0^2})/({2P})$ is what the minus component of the momentum $p_0$ would be if the mother parton were on shell. When $p_0^-$ does not equal this value, there is a deficit of minus momentum and the parton can only exist in this state for an interval in $x^+$ given by the inverse of the minus-momentum deficit. That is, the interval in $x^+$ between the vertex where the mother parton was created and where it decays can be estimated by
\begin{equation}
\Delta x^+ \sim \left[
p_0^- - \frac{\bm p_0^2 + m_0^2}{2P}
\right]^{-1}
\;.
\end{equation}
The interval in space-time, $\Delta x^\mu$, between the two vertices is approximately in the plus-direction, so the corresponding time interval is $\Delta t = \Delta x^0 \approx \Delta x^+/\sqrt 2$. That is,
\begin{equation}
\Delta t \sim \frac{1}{\sqrt 2}\left[
p_0^- - \frac{\bm p_0^2 + m_0^2}{2P}
\right]^{-1}
\;.
\end{equation}
This gives
\begin{equation}
e^{-t} \sim \frac{1}{\sqrt {Q_0^2}\,\Delta t}
\end{equation}
or
\begin{equation}
t \sim \log\!\left(\sqrt{Q_0^2}\,\Delta t\right)
\;.
\end{equation}
That is, the shower time at which a parton splits is an estimate of the logarithm of the coordinate time interval needed for the splitting, normalized by $\sqrt{Q_0^2}$ to make it dimensionless.

\section{Shower time for initial state splittings}
\label{sec:initialstatesplittings}

We now repeat the derivation of section~\ref{sec:finalstatesplittings}, this time for initial state splittings in collisions of two hadrons A and B. We consider splitting of a parton from hadron A. We define components of vectors using a reference frame in which $Q_0$ is predominantly in the time direction. We approximate the hadron momenta $p_\LA$ and $p_\LB$ to be lightlike, with $2\, p_\LA \cdot p_\LB = s$. We will use $(+,-,\perp)$ components of vectors and choose the axes so that $p_\LA$ has only a plus component and $p_\LB$ has only a minus component. In this frame, initial state partons from hadron A have large plus components, small transverse components, and very small minus components.

We start with an initial state parton with momentum
\begin{equation}
p_0 = \left(P,\,\frac{\bm p_0^2 + m_0^2 + v_0^2}{2P},\,\bm p_0 \right)
\;.
\end{equation}
Here $v_0^2 = p_0^2 - m_0^2$, the virtuality of the initial state parton, is negative. This parton splits (in backward evolution) into two partons, 1 and 2. Parton 1 is the new initial state parton, while parton 2 is radiated. Thus
\begin{equation}
p_0 = p_1 - p_2
\;.
\end{equation}
We define
\begin{equation}
\begin{split}
p_1 ={}& \left(\frac{1}{\z}\,P,\,\z\,
\frac{\bm p_1^2 + m_1^2 + v_1^2}{2P},\,\bm p_1 \right)
\;,
\\
p_2 ={}& \left(\frac{1-\z}{\z}\,P,\,
\frac{\z}{1-\z}\,\frac{\bm p_2^2 + m_2^2 + v_2^2}{2P},\, \bm p_2 \right)
\;.
\end{split}
\end{equation}
Here $v_1^2 \le 0$ and $v_2^2 \ge 0$.

We can easily find the virtuality $v_0^2$ by comparing the minus momentum of parton 0 to the difference of the minus momenta of partons 1 and 2:
\begin{equation}
\begin{split}
v_0^2
={}&
\z\, v_1^2
- \frac{\z}{1-\z}\,v_2^2
- \frac{1}{1-\z}\,
\left(
\bm p_2 -
(1-\z)\bm p_1
\right)^{\!2}
+ \z m_1^2 - \frac{\z}{1-\z}\,m_2^2 - m_0^2
\;.
\end{split}
\end{equation}
We note that we can approximate $v_0^2$ by its value for on-shell daughter partons,
\begin{equation}
\label{eq:Q0sqapproxIS}
v_0^2
\approx
-\frac{1}{1-\z}\,
\left(
\bm p_2
- (1-\z) \bm p_1
\right)^{\!2}
+ \z m_1^2 - \frac{\z}{1-\z}\,m_2^2 - m_0^2
\;,
\end{equation}
provided that
\begin{equation}
\begin{split}
\z\, |v_1^2| \ll{}& |v_0^2|
\;,
\hskip 1 cm
\frac{\z}{1-\z}\, v_2^2  \ll |v_0^2|
\;.
\end{split}
\end{equation}
That is
\begin{equation}
\label{eq:ISlambdaordering1}
\frac{|v_1^2|}{2\,p_1\cdot n} \ll \frac{|v_0^2|}{2\,p_0\cdot n}
\end{equation}
and
\begin{equation}
\label{eq:ISlambdaordering2}
\frac{v_2^2}{2\,p_2\cdot n} \ll  \frac{|v_0^2|}{2\,p_0\cdot n}
\;,
\end{equation}
where $n$ is the lightlike vector $n = (0,Q_0^-,\bm 0)$. 

For the parton ``2'' that was radiated into the final state, since $p_2^+ \gg p_2^\perp \gg p_2^-$, we have $2\,p_2\cdot n \approx 2\,p_2\cdot Q_0$. For the initial state partons ``0'' and ``1,'' this same approximation is valid. However, later formulas are nicer if we stick with $p_0\cdot n$ and $p_1\cdot n$ using the lightlike vector $n$. A useful notation for this is provided by writing
\begin{equation}
\begin{split}
p_0 \cdot n ={}& \eta_0\, p_\LA \cdot Q_0
\;,
\\
p_1 \cdot n ={}& \eta_1\, p_\LA \cdot Q_0
\;,
\end{split}
\end{equation}
where the momentum fractions $\eta_0$ and $\eta_1$ are defined by
\begin{equation}
\begin{split}
\eta_0 ={}& 2\,p_0\cdot p_\LB/s
\;,
\\
\eta_1 ={}& 2\,p_1\cdot p_\LB/s
\;.
\end{split}
\end{equation}
With this notation, we can write the conditions for neglecting $v_1^2$ and $v_2^2$ as
\begin{equation}
\begin{split}
\frac{|v_1^2|}{2\eta_1\, p_\LA\cdot Q_0} \ll{}& \frac{|v_0^2|}{2 \eta_0\,p_\LA\cdot Q_0}
\;,
\hskip 1 cm
\frac{v_2^2}{2\, p_2\cdot Q_0} \ll  \frac{|v_0^2|}{2 \eta_0\,\cdot Q_0}
\;.
\end{split}
\end{equation}

This leads us to define the shower time for an initial state splitting that is analogous to eq.~(\ref{eq:showertime}) for a final state splitting,
\begin{equation}
\begin{split}
\label{eq:initialstateLambda}
e^{-t_i} ={}& \frac{\Lambda_i^2}{Q_0^2} = \frac{p_i^2 - m_i^2}{2p_i\cdot Q_0} 
\hskip 1.4 cm {\rm final\ state\ parton}
\;,
\\
e^{-t_i} ={}& \frac{\Lambda_i^2}{Q_0^2} = \frac{|p_i^2 - m_i^2|}{2\eta_i\,p_\LA\cdot Q_0} 
\hskip 1cm {\rm initial\ state\ parton}
\;.
\end{split}
\end{equation}
with an analogous equation for an initial state parton from hadron B. Because at each splitting a small amount of momentum needs to be taken from the existing final state partons in the event, as described in section \ref{sec:momentumconservation} below, there is an ambiguity in how exactly to calculate $\eta_i$. We define $\eta_i$ to be the momentum fraction of the initial state parton as it existed when the parton was created, before it splits.

\section{Interference graphs}

\begin{figure}
\centerline{\includegraphics[width=7.0cm]{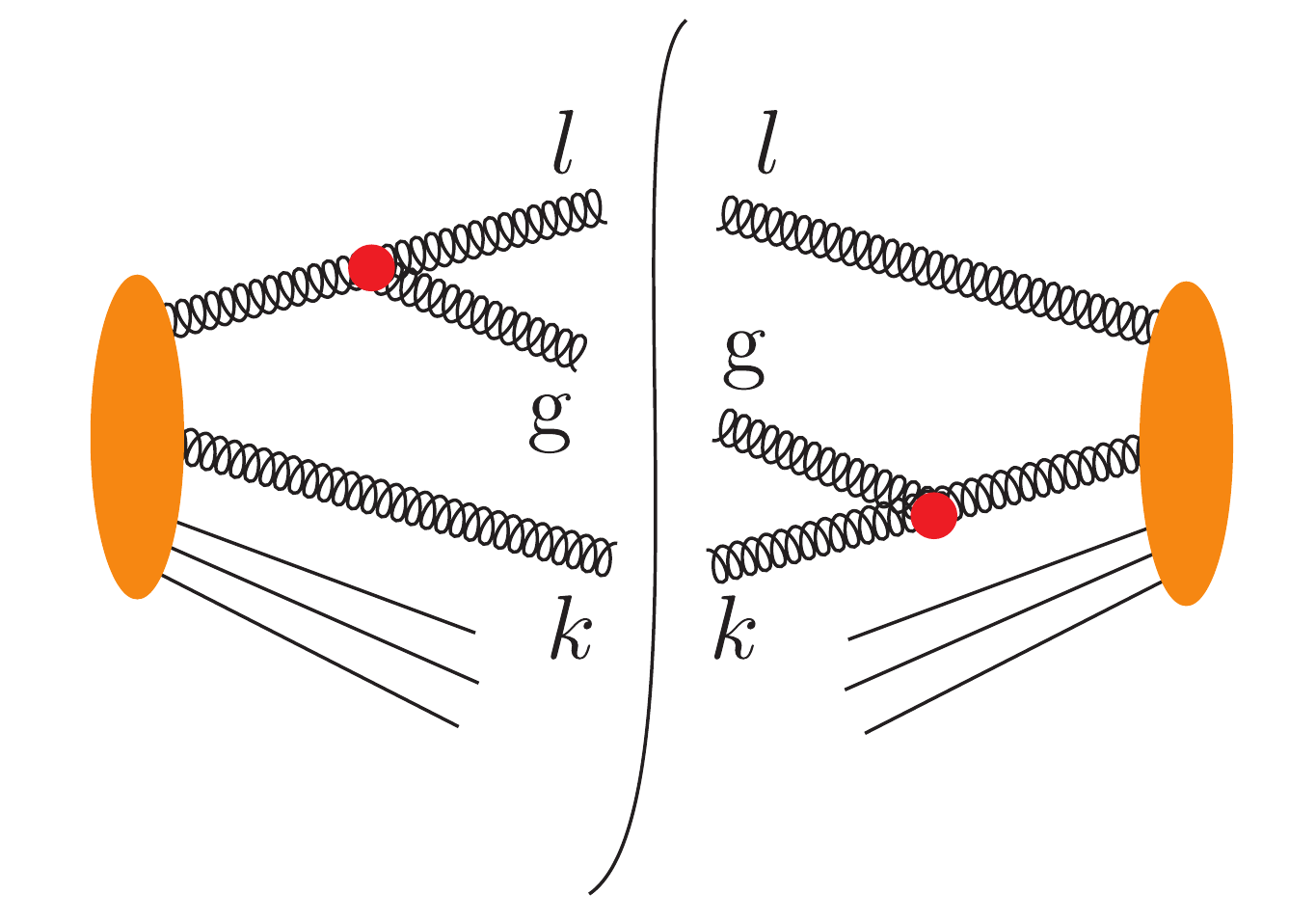}}
\caption{Emission of a soft gluon $\Lg$ from parton $l$ in ${\cal M}$ and from parton $k$ in ${\cal M}^*$.}
\label{fig:interference}
\end{figure}

A parton with label $l$ can emit a gluon, giving an amplitude ${\cal M}_l$ (in a physical gauge). We have, so far, analyzed the choice of ordering for squared amplitudes, $|{\cal M}_l|^2$. However, we need to account for quantum interference. A parton with label $k$, moving in a different direction, can emit the gluon, giving an amplitude ${\cal M}_k$. When the gluon is soft and its direction is not highly collinear with either parton $l$ or parton $k$, both processes are important and should be included in a parton shower. Thus we need to account for interference contributions ${\cal M}_l\, {\cal M}_k^*$ and ${\cal M}_k\, {\cal M}_l^*$, as illustrated in figure \ref{fig:interference}

The parton shower algorithm of \cite{deductor,NSI,NSII} takes such interference contributions into account, at least approximately. To understand how this works, consider emission of a gluon with momentum $\hat p_\Lg$ from parton $l$ with interference from emission from parton $k$. The momenta of the partons $l$ and $k$ are $p_l$ and $p_k$ before the emission and $\hat p_l$ and $\hat p_k$ after the emission. The probability for this emission in the limit that the gluon is soft, $\hat p_\Lg \to 0$, is proportional to the dipole splitting function
\begin{equation}
\label{eq:wlkdipole}
\overline w_{lk}^{\,\rm dipole}
= 4\pi\as\
\frac{
-(\hat p_{\Lg}\cdot \hat p_l\ \hat p_k
- \hat p_{\Lg}\cdot \hat p_k\ \hat p_l)^2}
{(\hat p_{\Lg}\cdot \hat p_k\ \hat p_{\Lg}\cdot \hat p_l)^2}
\;\;.
\end{equation}
This splitting function is described in more detail in sections~(5.3) and (5.5) of ref.~\cite{NScolor}. In eq.~(\ref{eq:wlkdipole}), partons $l$ and $k$ can have nonzero masses. The expression for  $\overline w_{lk}^{\,\rm dipole}$ is simpler in the massless limit, $\hat p_l^2 \to 0$ and $\hat p_k^2 \to 0$, where it becomes
\begin{equation}
\overline w_{lk}^{\,\rm dipole}
\to 4\pi\as\
\frac{
2\hat p_{k}\cdot \hat p_l}
{\hat p_{\Lg}\cdot \hat p_k\ \hat p_{\Lg}\cdot \hat p_l}
\;\;.
\end{equation}
This splitting function multiplies an appropriate matrix $C_{lk}$ in color space, with $C_{lk} = C_{kl}$. Eq.~(\ref{eq:wlkdipole}) includes all four diagrams for emission from either parton $l$ or parton $k$ in the amplitude and the conjugate amplitude, calculated in the limit $\hat p_{\Lg} = \lambda P_{\Lg}$ with $\lambda \to 0$.

We introduce a partitioning function $A'_{lk}$ with the properties that $A'_{lk} > 0$ and $A'_{lk} + A'_{kl} = 1$ as follows\footnote{The color factors $C_{lk}$ are the factors in square brackets in eq.~(5.7) of ref.~\cite{NScolor}. Equation (\ref{eq:insertAlk}) appeared as eq.~(5.8) of that paper, but there we inadvertently left out the color factors $C_{lk}$.}
\begin{equation}
\begin{split}
\label{eq:insertAlk}
\frac{1}{2}
\sum_l\sum_{k\ne l} 
\overline w_{lk}^{\rm dipole} C_{lk}
 ={}&
\frac{1}{2}
\sum_l\sum_{k\ne l} 
[A'_{lk} + A'_{kl}]
\overline w_{lk}^{\rm dipole}C_{lk}
\\={}&
\sum_l\sum_{k\ne l} 
A'_{lk}\overline w_{lk}^{\rm dipole}C_{lk}
\;\;.
\end{split}
\end{equation}
We define $A'_{lk}$ as in eq.~(7.12) of ref.~\cite{NSspin}:
\begin{equation}
\label{eq:Alkprime}
A'_{lk} 
=
\frac{\hat p_\Lg \cdot \hat p_k\ \hat p_l \cdot \hat Q}
{\hat p_\Lg \cdot \hat p_k\ \hat p_l \cdot \hat Q + \hat p_\Lg \cdot \hat p_l\ \hat p_k \cdot \hat Q}
\;.
\end{equation}
Here $\hat Q$ is the total momentum of the final state particles just after the splitting. We see
that $A'_{lk} \to 1$ and $A'_{kl} \to 0$ when $\hat p_\Lg$ becomes collinear with $\hat p_l$. We treat the term $A'_{lk}\overline w_{lk}^{\rm dipole}C_{lk}$ as primarily describing the emission of the gluon from parton $l$, with parton $k$ playing a passive role as a spectator. For instance, there is a momentum mapping that takes a small amount of momentum from the rest of the partons in the event and delivers it to the partons involved in the splitting so that they can be on shell both before and after the splitting. We use the momentum mapping associated with $p_l \to \hat p_l + \hat p_\Lg$ and simply ignore the small momentum transfer to parton $k$. More importantly for the topic of this paper, we define the shower time from the splitting $p_l \to \hat p_l + \hat p_\Lg$.

This seems rather crude. Can it be sensible? Note that the shower time associated with gluon emission from parton $l$ is $t = -\log(\Lambda_l^2/Q_0^2)$, where
\begin{equation}
\Lambda_l^2 \approx 
\frac{|(\hat p_l \pm \hat p_\Lg)^2 - m_l^2|}{2(\hat p_l \pm \hat p_\Lg)\cdot Q_0 }\,Q_0^2
=
\frac{2 \hat p_l \cdot \hat p_\Lg}{2(\hat p_l \pm \hat p_\Lg)\cdot Q_0 }\,Q_0^2
\;.
\end{equation}
Here the $+$ sign is for a final state splitting while the $-$ sign is for an initial state splitting. Of course, the same definition with  $l \to k$ applies for $\Lambda_k^2$. For small $\hat p_\Lg$, this becomes
\begin{equation}
\Lambda_l^2 \approx 
\frac{\hat p_l \cdot \hat p_\Lg}{\hat p_l \cdot Q_0 }\,Q_0^2
\;.
\end{equation}
Compare this to the same function using $\hat Q$ in place of $Q_0$:
\begin{equation}
\widetilde\Lambda_l^2 \approx 
\frac{\hat p_l \cdot \hat p_\Lg}{\hat p_l \cdot \hat Q }\,\hat Q^2
\;.
\end{equation}
We have
\begin{equation}
\widetilde\Lambda_l^2 \approx  \alpha_l \Lambda_l^2
\;,
\end{equation}
where
\begin{equation}
\alpha_l = \frac{\hat Q^2}{Q_0^2}\, \frac{\hat p_l \cdot Q_0}{\hat p_l \cdot \hat Q}
\;.
\end{equation}
Thus
\begin{equation}
\label{eq:Alkprime2}
A'_{lk} \approx \frac{\alpha_k \Lambda_k^2}{\alpha_l\Lambda_l^2 +  \alpha_k\Lambda_k^2}
\;.
\end{equation}
Note that only the ratio of $\alpha_l$ to $\alpha_k$, not their individual values, matters in $A'_{lk}$. Furthermore, $\alpha_l/\alpha_k$ is typically neither much larger than 1 nor much smaller than 1.

What happens? We can consider three cases.

First, one can have $\Lambda_l^2 \ll \Lambda_k^2$. Then $A'_{lk} \approx 1$ and $A'_{kl} \approx 0$, so the splitting is treated almost entirely as gluon emission from parton $l$, for which $\Lambda^2_l$ defines the shower time. Thus $\Lambda^2_l$ must be smaller than the $\Lambda^2$ of the previous splitting in the shower. In this case, $\Lambda^2_k$ is much larger than $\Lambda^2_l$ and may be larger than the $\Lambda^2$ of the previous splitting, so that the approximations needed to neglect $(\hat p_k \pm \hat p_\Lg)^2 - m_k^2$ in the previous splitting that produced parton $k$ may not be valid. However, in this case, if we examine the graphs that go into $\overline w_{lk}^{\rm dipole}$, using a physical gauge $\hat Q \cdot A = 0$, we find that the dominant graph is the one in which the soft gluon is emitted from parton $l$ both in the amplitude and in the conjugate amplitude. Graphs involving emission from parton $k$ are suppressed. Thus it does not matter if the approximations do not work well for emission from parton $k$.

Second, one can have $\Lambda_k^2 \ll \Lambda_l^2$. This is the same as the previous case but with $l \leftrightarrow k$.

In the third case, $\Lambda_k^2$ and $\Lambda_l^2$ are of a similar size. In this case, we really do have substantial quantum interference in a physical gauge. The shower algorithm sometimes uses $\Lambda_l^2$ to define the shower time and sometimes uses $\Lambda_k^2$. However, it does not much matter which $\Lambda^2$ value is used because the two are of similar size.

One could, of course, use $Q_0$ in place of $\hat Q$ in the definition of $A'_{lk}$. Then the argument given above would be simpler. Our only reason for not doing that is that the code for generating parton splittings is somewhat simpler with $A'_{lk}$ defined using $\hat Q$.

\section{Connection to $k_\LT$ ordering}
\label{sec:kT}

\begin{figure}
\centerline{\includegraphics[width=14.0cm]{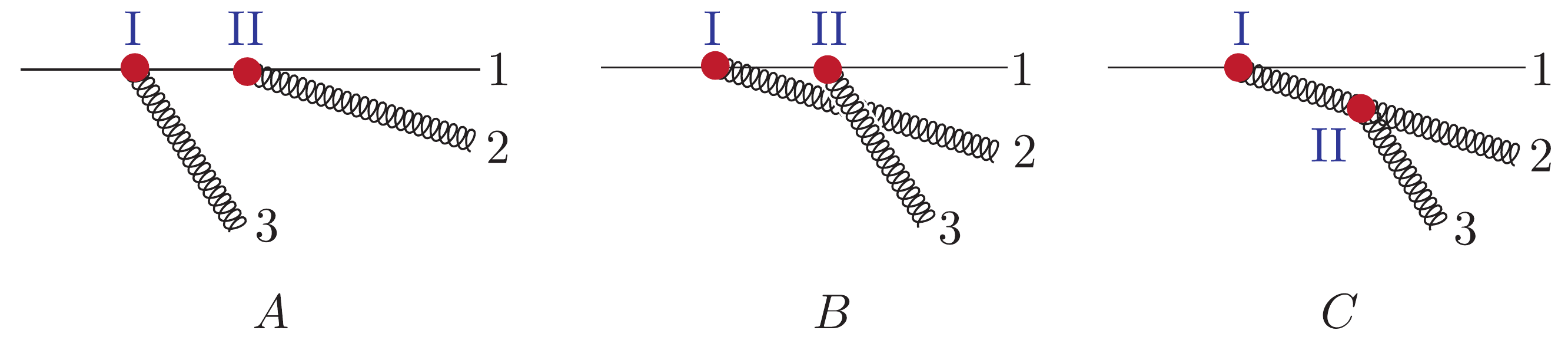}}
\caption{Three possible shower histories.}
\label{fig:HistoriesABC}
\end{figure}

To see if the choice of shower time makes a difference, we consider an example in which a final state quark splits into a quark, labelled 1, and two gluons, labelled 2 and 3. For simplicity, in this section we take the quark to be massless. There are three possible shower histories, illustrated in figure~\ref{fig:HistoriesABC}. Which histories can be generated in a parton shower depends on the momenta $p_1$, $p_2$ and $p_3$ and on the choice of the shower time variable.

\subsection{Kinematics}

Consider a final state splitting of a massless mother parton to massless daughter partons, $i$ and $j$. We think of the partons as part of a jet of momentum $p_\LJ$, approximated as lightlike.  We define $(+,-,\perp)$  components of vectors so that the plus direction is in the direction of $p_\LJ$:
\begin{equation}
p_{\LJ} = (P,\,0,\,\bm 0)
\;.
\end{equation}
As in eq.~(\ref{eq:p12components}), we take the $(+,-,\perp)$ components of the momenta of partons to be
\begin{equation}
\begin{split}
\label{eq:pijcomponents}
p_i ={}& \left(x_i P,\,\frac{\bm p_i^2}{2x_i P},\,\bm p_i \right)
\;,
\\
p_j ={}& \left(x_j P,\,\frac{\bm p_j^2}{2x_j P},\,\bm p_j \right)
\;.
\end{split}
\end{equation}
Partons $i$ and $j$ may have daughter partons and thus have nonzero virtualities, but we are neglecting their virtualities here. 

For the splitting of the mother parton into partons $i$ and $j$, we can define a squared transverse momentum $k_\LT^2$ by
\begin{equation}
k_\LT^2 = \frac{1}{(x_i + x_j)^2}\,(x_j \bm p_i - x_i\, \bm p_j)^2
\;\;.
\end{equation}
One justifies the name $k_\LT^2$ by noting that if $\bm p_i + \bm p_j = 0$, then $k_\LT^2 = \bm p_i^2$. Furthermore, if we boost the momenta using a null plane boost $\bm p_i \to \bm p_i + x_i \bm v$ and $\bm p_j \to \bm p_j + x_j \bm v$,  we can reach a frame in which $\bm p_i + \bm p_j = 0$ while leaving $k_\LT^2$ invariant. In a $k_\LT$ ordered shower, one orders splittings so that $k_\LT^2$ for a daughter splitting is smaller than $k_\LT^2$ for the mother splitting.

The virtuality of the mother parton is related to $k_\LT^2$ by
\begin{equation}
(p_i + p_j)^2 = \frac{(x_i + x_j)^2}{x_i x_j}\,k_\LT^2
\;.
\end{equation}
Ordering according to shower time as defined in eq.~(\ref{eq:showertime}) amounts to ordering in the variable
\begin{equation}
\Lambda^2 = \frac{(p_i + p_j)^2}{x_i + x_j}\ \frac{Q_0^2}{2 p_\LJ \cdot Q_0}
\;.
\end{equation}
The two ordering variables are related by
\begin{equation}
\Lambda^2 = \frac{x_i + x_j}{x_i x_j}\,k_\LT^2\ \frac{Q_0^2}{2 p_\LJ \cdot Q_0}
\;.
\end{equation}

Consider now the shower history $A$ in figure~\ref{fig:HistoriesABC}. We take $P$ to represent the plus-momentum of the mother parton, so that $x_1 + x_2 + x_3 = 1$. There are two splittings, which we can call I and II. The $k_\LT$ values for the two splittings are
\begin{equation}
\begin{split}
k_\LT^2(\LI) ={}& \big( (x_1 + x_2) \bm p_3 - x_3 (\bm p_1 + \bm p_2) \big)^2
\;,
\\
k_\LT^2(\LI\LI)  ={}& \frac{1}{(x_1 + x_2)^2}\,(x_1 \bm p_2 - x_2\, \bm p_1)^2
\;.
\end{split}
\end{equation}
It will prove useful to exchange the transverse momentum variables for two dimensional angular variables, defined by
\begin{equation}
\bm \theta_i = \frac{\sqrt 2}{P}\, \frac{\bm p_i}{x_i}
\;.
\end{equation}
We define differences in angles by
\begin{equation}
\bm \theta_{ij} = \bm \theta_i - \bm \theta_j
\;.
\end{equation}

Using these angular variables, the $k_\LT^2$ values for the two splittings are
\begin{equation}
\begin{split}
\label{eq:kTordering}
k_\LT^2(\LI) ={}&  \frac{(p_J\cdot Q_0)^2}{Q_0^2}\ 
x_3^2
\big( x_1 \bm \theta_{31} + x_2 \bm \theta_{32} \big)^2
\;,
\\
k_\LT^2(\LI\LI)  ={}& \frac{(p_J\cdot Q_0)^2}{Q_0^2}\ 
\frac{(x_1 x_2)^2}{(x_1 + x_2)^2}\,\bm \theta_{12}^2
\;.
\end{split}
\end{equation}
The corresponding $\Lambda^2$ values for ordering according to eq.~(\ref{eq:showertime}) are
\begin{equation}
\begin{split}
\label{eq:Lambdaordering}
\Lambda^2(\LI) ={}&  \frac{p_J\cdot Q_0}{2}\,
\frac{x_3}{x_1 + x_2}
\big( x_1 \bm \theta_{31} + x_2 \bm \theta_{32} \big)^2
\;,
\\
\Lambda^2(\LI\LI)  ={}&  \frac{p_J\cdot Q_0}{2}\,
\frac{x_1 x_2}{x_1 + x_2}\,\bm \theta_{12}^2
\;.
\end{split}
\end{equation}
Thus history A is a valid shower history for partons 1, 2, and 3 according to $k_\LT$ ordering if $k_\LT^2(\LI) > k_\LT^2(\LI\LI)$ and it is a valid history according to $\Lambda$ ordering if $\Lambda^2(\LI) > \Lambda^2(\LI\LI)$. The same analysis applies to histories B and C. We simply have to permute the labels 1,2,3 in eqs.~(\ref{eq:kTordering}) and (\ref{eq:Lambdaordering}).

\subsection{A specific case}

Mostly, $\Lambda$ ordering and $k_\LT$ ordering are equivalent for splittings to make three partons. Here is one case where they differ. Suppose that $x_3 \ll x_2 \ll 1$. Then we take $x_1 \approx 1$ since $x_1 + x_2 + x_3 = 1$. Define $\bm\theta = (\bm\theta_{31} + \bm\theta_{32})/2$ and suppose that $\bm\theta^2_{12} \ll \bm\theta^2$. Then $\bm\theta^2_{31} \approx \bm\theta^2_{32} \approx \bm\theta^2$. Under these circumstances, for history A we have
\begin{equation}
\begin{split}
\label{eq:kTorderingA}
k_\LT^2(\LI) ={}&  \frac{(p_J\cdot Q_0)^2}{Q_0^2}\  
x_3^2\,
\bm \theta^2
\;,
\\
k_\LT^2(\LI\LI)  ={}& \frac{(p_J\cdot Q_0)^2}{Q_0^2}\ 
x_2^2\,\bm \theta_{12}^2
\end{split}
\end{equation}
and
\begin{equation}
\begin{split}
\label{eq:LambdaorderingA}
\Lambda^2(\LI) ={}&   \frac{p_J\cdot Q_0}{2}\   
x_3\,
\bm \theta^2
\;,
\\
\Lambda^2(\LI\LI)  ={}& \frac{p_J\cdot Q_0}{2}\  
x_2\,\bm \theta_{12}^2
\;.
\end{split}
\end{equation}
Thus history A is allowed for $k_\LT$ ordering if
\begin{equation}
\frac{\bm \theta_{12}^2}{\bm \theta^2} <  \frac{x_3^2}{x_2^2}
\end{equation}
while history A is allowed for $\Lambda$ ordering if
\begin{equation}
\frac{\bm \theta_{12}^2}{\bm \theta^2} <  \frac{x_3}{x_2}
\;.
\end{equation}

Suppose that
\begin{equation}
\label{eq:specialregion}
\frac{x_3^2}{x_2^2} \ll
\frac{\bm \theta_{12}^2}{\bm \theta^2} \ll  \frac{x_3}{x_2}
\;.
\end{equation}
Then history A is allowed for $\Lambda$ ordering. However, history A is not generated with $k_\LT$ ordering. With the same sort of analysis, we find instead that with $k_\LT$ ordering we generate histories B and C, while with $\Lambda$ ordering histories B and C are forbidden.

How can we interpret this result? The region in which $x_3 \ll x_2 \ll 1$ and $\bm \theta_{12}^2 \ll \bm\theta^2 \ll 1$ with $\bm\theta_{12}^2/\bm \theta^2$ limited by eq.~(\ref{eq:specialregion}) is important. It can generate four large logarithms, two from two angle integrations and two from two momentum fraction integrations. The analysis of section~\ref{sec:finalstatesplittings} indicates that both denominators in the Feynman graph corresponding to history A become independently small in this region, so that the square of graph A (in a physical gauge) gives a leading contribution. However, this does not work for graphs B and C. For instance, in graph B, the $(p_1 + p_3)^2$ is so large that it dominates the denominator proportional to $(p_1 + p_2 + p_3)^2$ as represented in eq.~(\ref{eq:Q0sq}). Then the needed sensitivity to $p_2$ is absent. Thus the leading Feynman graph in the amplitude for $0 \to 1 + 2 + 3$ is that corresponding to history A.

What happens, then, if one uses a $k_\LT$ ordered shower, which generates the shower using histories B and C but not A? Surprisingly, we get the right answer. There are two reasons for this. 

The first reason concerns the propagator denominator corresponding to splitting I in graphs B and C. Consider graph B. The momentum $p_1 + p_3$ is really significantly off shell, considering that the corresponding parton subsequently splits to partons 1 and 3. However, when we generate it, we treat it as exactly on shell. Thus when we generate splitting I in history B we use essentially the same propagator denominator as when we generate splitting II in history A. (However, depending on how the shower algorithm treats momentum conservation, the angle between partons 1 and 2 may be adjusted significantly to account for the recoil from parton 3.)

The second reason concerns the amplitude for splitting II in graph B, in which parton 3 is emitted. This is an emission of a soft, wide angle gluon. The amplitude can be approximated using the eikonal approximation,
\begin{equation}
{\cal A} = \frac{g\, n\cdot \epsilon_3}{n\cdot p_3}
\;,
\end{equation}
where $n$ is a lightlike vector in the direction of the jet, $n = (1,0,0,0)$. Here $\epsilon_3$ is the polarization vector of gluon 3. The same approximation applies to splitting II in graph C. Finally, this same approximation applies to splitting I in graph A. Thus the kinematic factor describing the emission of parton 3 is the same in graphs A, B, and C. The three graphs have different color factors, which we may call ${\cal C}_A$, ${\cal C}_B$, and ${\cal C}_C$, respectively. Color invariance implies that
\begin{equation}
{\cal C}_B + {\cal C}_C = {\cal C}_A
\;.
\end{equation}

Putting these two arguments together, we see that when we add the parton shower approximations corresponding to histories B and C, we get the parton shower approximation corresponding to history A. That is, even though using a $k_\LT$ ordered shower generates the given three parton configuration according to histories B and C instead of A, the result is approximately the same.

An analogous argument shows that one can also use angular ordering, in which history A applies whenever $\bm\theta_{12} \ll \bm\theta$. This, of course, is the basic physics argument behind angle ordered showers \cite{angleorder}.

\section{Consequences for initial state splittings}
\label{sec:InitialState}

We now turn to the effect of using $\Lambda$ ordering in the initial state shower. We consider the case in which all of the partons are gluons. We seek insight into what initial state emissions are allowed by $\Lambda$ ordering.

\subsection{Kinematics}

\begin{figure}
\centerline{\includegraphics[width=6.0cm]{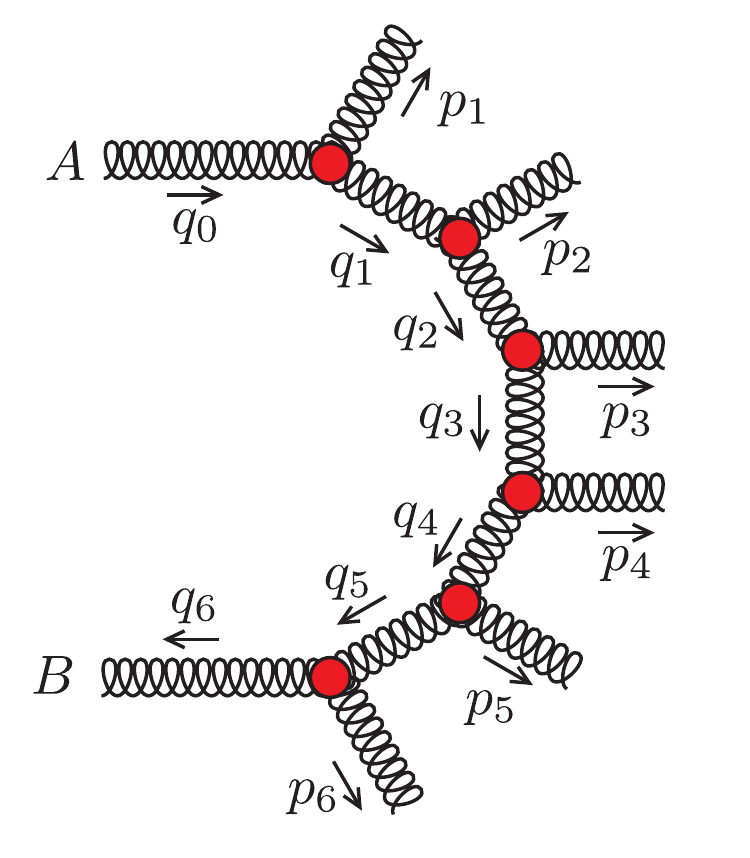}}
\caption{Initial state shower history showing initial state gluons with momenta $\{q_0,\dots,q_N\}$ and emitted gluons with momenta $\{p_1,\dots,p_N\}$ with $N = 6$.}
\label{fig:bfkldiagram}
\end{figure}

The kinematics for the initial state shower is illustrated in figure~\ref{fig:bfkldiagram}. We have a sequence of initial state parton momenta, which we denote by $$\{q_0,\, q_1, \dots,\, q_{h-1},\, q_h,\, q_{h+1}, \dots,\, q_{N-1},\, q_N\}.$$  Gluon 0 is the constituent of hadron A at the very end of the shower as we move from hard to soft interactions and gluon $N$ is the constituent of hadron B at the soft end of the initial state shower. The hard interaction is associated with the exchange of a gluon with momentum $q_h$. 

As in section \ref{sec:initialstatesplittings}, we treat the hadron momenta as being lightlike: $p_\LA^2 = p_\LB^2 = 0$. Then $s = (p_\LA + p_\LB)^2$ is given by $s = 2 p_\LA \cdot p_\LB$. We use a reference frame in which $p_\LA$ has only a plus component, $p_\LA^+ > 0$, with $p_\LA^- = 0$ and $p_\LA^\perp = 0$.  Similarly, $p_\LB$ has only a minus component, $p_\LB^- > 0$, with $p_\LB^+ = 0$ and $p_\LB^\perp = 0$. 

We treat all of the momenta $p_i$ of emitted gluons as being on shell: $p_i^2 = 0$. We take initial state gluon 0 to be on shell and collinear with hadron A: $q_0 = \eta_0 p_\LA$.  We take initial state gluon $N$ to be on shell and collinear with hadron B: $-q_N = \xi_N p_\LB$.  Momentum is conserved at each vertex: $q_{i-1} = q_i + p_i$. Then the $q_i$ for $0<i<N$ are spacelike and generally have non-zero transverse components. 

This is not the way that the shower is generated. In actually generating the initial state shower, at each splitting stage, we treat the new initial state gluon as being on shell with zero transverse momentum. This involves some approximations, which we outline in section \ref{sec:momentumconservation}. In this section, we ignore these complications and simply use spacelike initial state partons with non-zero transverse components and exact momentum conservation.

For each initial state gluon, decompose $q_i$ according to 
\begin{equation}
q_i = \eta_i p_A - \xi_i p_B + q_i^\perp
\;,
\end{equation}
where $q_i^\perp \cdot p_\LA = q_i^\perp \cdot p_\LB = 0$. Because $p_i \cdot p_\LA > 0$ and $p_i \cdot p_\LB > 0$, we have $\eta_i > \eta_{i+1}$ and $\xi_i > \xi_{i-1}$. Also $\xi_0 = 0$ and $\eta_N = 0$. That is, the momentum fractions $\eta_i$ decrease to zero as we move from hadron A to hadron B and the momentum fractions $\xi_i$ decrease to zero as we move from hadron B to hadron A. The initial state gluon momenta for $0 < i < N$ are all spacelike:
\begin{equation}
\label{eq:qisquare}
q_i^2 = - \eta_i \xi_i \,s - \bm q_i^2  < 0
\;.
\end{equation}
Here and in the following, we denote the two components of $q_i^\perp$ with a boldface symbol $\bm q_i$, using the euclidian inner product in the transverse space so that $\bm q_i^2 > 0$.

\subsection{Evolution of the gluon virtualities}

There is an instructive relation between $q_i^2$ and $q_{i-1}^2$. To derive this relation, we start with the identity
\begin{equation}
\label{eq:identity0}
0 < [(\eta_{i-1} - \eta_i) \bm q_i - \eta_i \bm p_i]^2
\;.
\end{equation}
Using $\bm q_{i-1} = \bm q_{i} + \bm p_{i}$, this identity is equivalent to
\begin{equation}
\label{eq:identity1}
\frac{\bm q_{i-1}^2}{\eta_{i-1}} < 
\frac{\bm q_{i}^2}{\eta_{i}}
+ \frac{\bm p_{i}^2}{\eta_{i-1} - \eta_{i}}
\;.
\end{equation}
The condition that $p_i$ is lightlike is
\begin{equation}
\label{eq:ponshell}
\bm p_i^2 = (\eta_{i-1} - \eta_i) (\xi_i - \xi_{i-1}) \,s 
\;.
\end{equation}
Thus eq.~(\ref{eq:identity1}) is equivalent to
\begin{equation}
\label{eq:identity2}
\frac{\bm q_{i-1}^2 + \eta_{i-1}\xi_{i-1}\,s}{\eta_{i-1}} < 
\frac{\bm q_{i}^2 + \eta_{i}\xi_{i}\,s}{\eta_{i}}
\;.
\end{equation}
Using eq.~(\ref{eq:qisquare}), this is
\begin{equation}
\label{eq:identity3}
\frac{- q_{i-1}^2}{\eta_{i-1}} < 
\frac{- q_{i}^2}{\eta_{i}}
\;.
\end{equation}
This is the $\Lambda$ ordering condition (\ref{eq:ISlambdaordering1}) for an initial state shower starting at the hard interaction and moving toward hadron A except that ``$\ll$'' is replaced by ``$<$.'' However, the actual shower generation uses ``$<$.''

With an analogous derivation, we have
\begin{equation}
\label{eq:identity5}
\frac{- q_{i+1}^2}{\xi_{i+1}} < 
\frac{- q_{i}^2}{\xi_{i}}
\;.
\end{equation}
This is the $\Lambda$ ordering condition for an initial state shower starting at the hard interaction and moving toward hadron B. 

We thus see that shower splittings that violate $\Lambda$ ordering in the form (\ref{eq:ISlambdaordering1}) are impossible if we use exact kinematics in the initial state shower. However, we do not use exact kinematics. Instead, we approximate initial state partons as being exactly on shell, then adjust them to have spacelike momenta by taking momentum from elsewhere in the event, as described in the following subsection. When we make these approximations, it is important to impose $\Lambda$ ordering so that we do not generate splittings that invalidate the approximations.

\subsection{Momentum conservation}
\label{sec:momentumconservation}

The initial state shower does not appear all at once as depicted in figure \ref{fig:bfkldiagram}. Rather, it is generated step by step, starting from the hard interaction. To understand the relationship between the final result depicted in figure \ref{fig:bfkldiagram} and the steps in generating the shower, let us suppose that the shower has been partly generated. At the current stage of shower generation, the latest parton to be generated on the hadron A side is parton $i$, with $i < h$. The latest parton to be generated on the hadron B side is parton $j$, with $j > h$. We work in a reference frame, ``frame 0,'' in which these partons have zero transverse momenta. We approximate them as being on shell. That is, we are neglecting their virtualities as having negligible effect on the calculation of all harder interactions. Thus partons $i$ and $j$ have momentum components
\begin{equation}
\begin{array}{rll}
q_i ={}& (\eta_i p_\LA^+,\,0,\,\bm 0)  &\hskip 2 cm \text{frame 0}
\;,
\\
-q_j ={}& (0,\,\xi_j p_\LB^-,\,\bm 0)  &\hskip 2 cm \text{frame 0}
\;.
\end{array}
\end{equation}

Now we let parton $i$ split. Unfortunately, a parton with zero virtuality cannot split. Thus we need to change $q_i$ to a new momentum $\tilde q_i$ with 
\begin{equation}
\label{eq:tildeqi}
\tilde q_i = \left(\tilde\eta_i p_\LA^+,\,
\frac{\tilde q_i^2}{2 \tilde\eta_i p_\LA^+},\,
\bm 0\right)  
\hskip 2 cm \text{frame 0}
\;.
\end{equation}
We want the shower to exactly conserve momentum, so we will have to take the needed momentum $\tilde q_i - q_i$ from elsewhere in the event. We get the needed momentum by applying a small Lorentz transformation $\Lambda_1(\omega)$ to every final state parton that exists at this stage of the evolution,
\begin{equation}
p_k \to \tilde p_k = \Lambda_1(\omega) p_k
\hskip 2 cm k = i+1,\dots,j
\;.
\end{equation}
Then if we define
\begin{equation}
Q_{ij} = \sum_{k = i+1}^j p_k
\;,
\end{equation}
we have
\begin{equation}
Q_{ij} \to \widetilde Q_{ij} = \Lambda_1(\omega) Q_{ij}
\end{equation}
The original $Q_{ij}$ obeys
\begin{equation}
q_i = Q_{ij} + q_j
\;.
\end{equation}
If we now use momentum conservation with the shifted final state momenta, we have a modified $q_i$,
\begin{equation}
\label{eq:Qij}
\tilde q_i = \tilde Q_{ij} + q_j
\;.
\end{equation}
That is
\begin{equation}
\begin{split}
\tilde q_i ={}& \Lambda(\omega)Q_{ij} + q_j
\\
={}& \Lambda_1(\omega)[q_i - q_j] + q_j
\\
={}& \Lambda_1(\omega)q_i + (1-\Lambda_1(\omega))q_j
\;.
\end{split}
\end{equation}

We choose the Lorentz transformation $\Lambda_1(\omega)$ to be a boost in the $z$-direction with boost angle $\omega$. Then the components of $\tilde q_i$ are
\begin{equation}
\tilde q_i =
\left(e^\omega\eta_i p_\LA^+,\,
-(1-e^{-\omega})\xi_j p_\LB^-,\,
\bm 0 \right)
\;.
\end{equation}
With this, we obtain the square of $\tilde q_i$ as a function of $\omega$,
\begin{equation}
\begin{split}
- \tilde q_i^2
={}&
(e^\omega - 1) 2\eta_i \xi_j p_\LA\cdot p_\LB
\\
={}& (e^\omega - 1) Q_{ij}^2
\;,
\end{split}
\end{equation}
where we have used eq.~(\ref{eq:Qij}). Thus
\begin{equation}
e^\omega = 1 + \frac{-\tilde q_i^2}{Q_{ij}^2}
\;.
\end{equation}
Since the virtuality of any exchanged gluon is typically much smaller than the square of the momenta of all of the final state particles created between $i$ and $j$, we need only a small boost: $\omega \ll 1$. We see that the momentum fraction $\tilde \eta_i$ in eq.~(\ref{eq:tildeqi}) is slightly different from $\eta_i$:
\begin{equation}
\tilde \eta_i = e^\omega \eta_i
\;.
\end{equation}

Now that gluon $i$ has a nonzero virtuality, it can split in backward evolution to a new initial state gluon with momentum $q_{i-1}$ and a final state gluon with momentum $p_i$, with
\begin{equation}
\label{eq:momentumconservation}
q_{i-1} - p_i = \tilde q_i
\;,
\end{equation}
where $\tilde q_i$ is given by eq.~(\ref{eq:tildeqi}). We take $q_{i-1}$ and $p_i$ to be lightlike with components
\begin{equation}
\begin{array}{rll}
q_{i-1} ={}& \displaystyle{\left(\eta_{i-1} p_\LA^+,\,
\frac{\bm k_i^2}{2 \tilde z_i^2\eta_{i-1} p_\LA^+},\,
\frac{1}{\tilde z_i}\,\bm k_i\right) } 
&\hskip 2 cm \text{frame 0}
\;,
\\
p_i ={}& \displaystyle{\left((1-\tilde z_i)\eta_{i-1} p_\LA^+,\,
\frac{\bm k_i^2}{2 \tilde z_i^2(1-\tilde z_i)\eta_{i-1} p_\LA^+},\,
\frac{1}{\tilde z_i}\,\bm k_i\right) }   
&\hskip 2 cm \text{frame 0}
\;.
\end{array}
\end{equation}
Here $q_{i-1}$ has a new momentum fraction $\eta_{i-1}$. We define
\begin{equation}
\label{eq:tildezdef}
\tilde z_i = \frac{\tilde \eta_i}{\eta_{i-1}}
\end{equation}
and assign a momentum fraction $(1-\tilde z_i)\eta_{i-1}$ to the emitted gluon. Then the plus component of momentum is conserved according to eq.~(\ref{eq:momentumconservation}). We assign transverse momentum $\bm k_i/\tilde z_i$ to both $q_{i-1}$ and $p_i$. Then the transverse components of momentum are also conserved. With a small calculation, we see that the minus component of momentum is conserved as long as $\bm k_i$ is chosen so that
\begin{equation}
-\tilde q_i^2 = \frac{\bm k_i^2}{1-\tilde z_i}
\;.
\end{equation}

We need one more step. We should change to a new reference frame, ``frame 1,'' so that the transverse components of $q_{i-1}$ are zero. This is simple with a null-plane boost $\Lambda_2(\bm v)$ such that a vector with components $P$ in frame 0 has components $P' =\Lambda_2(\bm v) P$ in frame 1, with
\begin{equation}
\label{eq:nullplaneboost}
P' = \left(
P^+,\, P^- + \frac{1}{2}\,P^+ \bm v^2 +  \bm P\cdot \bm v,\, 
\bm P + P^+ \bm v
\right)
\;.
\end{equation}
If we choose
\begin{equation}
\bm v = -\frac{1}{\tilde\eta_i p_\LA^+ }\, \bm k_i
\;,
\end{equation}
we have
\begin{equation}
\label{eq:newmomenta}
\begin{array}{rll}
q_{i-1} ={}& \Big(\eta_{i-1} p_\LA^+,\,
0,\,
\bm 0\Big)  \phantom{\bigg|}
&\hskip 2 cm \text{frame 1}
\;,
\\
p_i ={}& \displaystyle{\left((1-\tilde z_i)\eta_{i-1} p_\LA^+,\,
\frac{\bm k_i^2}{2 (1-\tilde z_i)\eta_{i-1} p_\LA^+},\,
\bm k_i\right) }   \phantom{\bigg|}
&\hskip 2 cm \text{frame 1}
\;,
\\
\tilde q_i ={}& \displaystyle{\left(\tilde\eta_i p_\LA^+,\,
\frac{\tilde q_i^2}{2 \eta_{i-1} p_\LA^+},\,
- \bm k_i\right) }   \phantom{\bigg|}
&\hskip 2 cm \text{frame 1}
\;,
\\
-q_j ={}& \Big(0,\,\xi_j p_B^-,\,\bm 0\Big)\phantom{\bigg|}
&\hskip 2 cm \text{frame 1}
\;.
\end{array}
\end{equation}
The components of the final state momenta $\tilde p_i$ for $i = i+1,\dots,j$ change according to the boost defined by eq.~(\ref{eq:nullplaneboost}). However, the boost does not change the components of a vector that lies entirely in the minus direction, so the components of $q_j$ are not changed.

The practical effect of this is that $q_{i-1}$ and $p_i$ are given by eq.~(\ref{eq:newmomenta}), while the components of the momenta $p_k$ of final state gluons are modified by the net Lorentz transformation $\Lambda_2(\bm v)\Lambda_1(\omega)$. This analysis has been for massless partons. We state the needed Lorentz transformation for the general case of massive partons in appendix \ref{sec:Appendix}. 

The important point of this subsection is not the precise form of the Lorentz transformations, but rather the idea that, in generating an initial state splitting, we first take a small amount of momentum from the final state that allows us to conserve momentum in the splitting, then we change reference frames so that the new initial state parton has zero transverse momentum. Thus the momenta of final state partons shift at each emission so as to recoil against the transverse momentum of a newly emitted parton.

\subsection{Dynamical regimes}

Now, consider possible dynamical regimes for the initial state shower of figure~\ref{fig:bfkldiagram}. One regime is the standard one with soft or collinear gluons emitted from the initial state gluons. For the hadron A side of the shower, the momentum fractions $z_i = \eta_{i}/\eta_{i-1}$ are either close to 1 for a soft gluon emission or finite, close to neither 0 or 1, for a collinear splitting. Then all of the $\eta_i$ are roughly the same size. Then eq.~(\ref{eq:identity3}) implies that the virtualities $-q_i^2$ decrease as we move from the hard interaction toward hadron A. If the $z_i$ are not close to 1, the splitting transverse momenta,
\begin{equation}
\label{eq:kidef}
\bm k_i = (1-z_i)\bm q_{i-1} - \bm p_i 
\end{equation}
also decrease since, from eq.~(\ref{eq:Q0sqapproxIS}), 
\begin{equation}
\bm k_i^2 \approx (1-z_i) (-q_i^2)
\;.
\end{equation}

There is another regime, which one can call the cut pomeron regime. All of the $\bm k_i$ can be of roughly the same size, but the $\z_i$ can be small: $\eta_{i-1} \gg \eta_i$. In this regime, also $\xi_{i+1} \gg \xi_i$. The cut pomeron regime can be important when $s = 2 p_\LA\cdot p_\LB$ is much larger that the $\bm k_i^2$. This regime has been extensively studied \cite{Fadin:1975cb, Kuraev:1976ge, Kuraev:1977fs, Balitsky:1978ic} and there are computer codes that generate events in the cut pomeron regime \cite{Schmidt, OrrStirling, CASCADE1, CASCADE2, HEJ}. Of course, there are possible regimes intermediate between the standard regime and the cut pomeron regime, but we content ourselves here with just these two idealized cases.

Since neither transverse momenta nor virtualities need to decrease during shower generation if splitting momentum fractions $\z_i$ are small, \textsc{Deductor} can generate events in the cut pomeron regime. However, there is one issue that we need to consider. We start with a hard interaction, but in figure~\ref{fig:bfkldiagram}, which is the hard interaction? It seems that we can start anywhere. We adopt the following procedure. Of all the initial state propagators in figure~\ref{fig:bfkldiagram}, one has the largest $|q_i^2|$. We take that to be the hard interaction, labelled $h$. Then we develop initial state showers on both the A side and the B side of the hard interaction. We want the gluon with $i = h$ to be the one with the largest virtuality, so we require 
\begin{equation}
\label{eq:pomeronkTcut}
|q_i^2| < |q_h^2|
\end{equation}
in the initial state shower splittings.  In figure~\ref{fig:bfkldiagram} as drawn, $q_h^2$ is the invariant $\hat t$ of the parton-parton scattering that constitutes the hard interaction. However, in generating the hard interaction, there is no distinction between the two final state partons. If we exchange them, $q_h^2$ is the invariant $\hat u$. Thus we define
\begin{equation}
|q_h^2| = \min(|\hat t|, |\hat u|)
\;.
\end{equation}

Note that applying the cut eq.~(\ref{eq:pomeronkTcut}) is by no means the same as imposing $k_\LT$ ordering or virtuality ordering on the initial state shower. No $|q_i^2|$ can be larger than $|q_h^2|$, but as we move from the hard interaction towards hadron A or towards hadron B, the $|q_i^2|$ can decrease, then increase again. Thus we can have a gluon emitted with high transverse momentum and large negative rapidity, another gluon emitted with high transverse momentum and large positive rapidity, and several gluons emitted with smaller transverse momenta and intermediate rapidities.

Although \textsc{Deductor} does generate events in the cut pomeron regime, it was not designed with this regime in mind. To adapt it for cut pomeron physics, one would need to incorporate suitable Sudakov factors for each propagator in figure \ref{fig:bfkl}. These Sudakov factors should represent unresolved real radiation together with virtual radiation. In the current version of \textsc{Deductor}, there is no Sudakov factor associated with the hard interaction and the factors associated with initial state gluon emissions are merely those that keep the hard scattering cross section from being changed by shower evolution. It would be fairly straightforward to create extra Sudakov factors as weights that would differ from 1 in the event of large rapidity separations. Investigation of what extra Sudakov factors to incorporate remains as a topic for future investigation.

\subsection{A numerical investigation}

In order to understand the effect in \textsc{Deductor} of initial state parton splittings with $\z \ll 1$ and of very forward gluon scattering in the hard interaction, we examine an observable that was calculated in ref.~\cite{HEJ} using the generator \textsc{High Energy Jets}. We examine multijet production in 7 TeV proton-proton collisions. The jets are defined using the anti-$k_\LT$ algorithm \cite{antikT} with R = 0.6, found with the aid of \textsc{FastJet} \cite{FastJet}. Only jets with $|\bm p_{i\perp}| > 30 \GeV$ and rapidity $|y_j| < 4.5$ are considered. Define $\Delta y = y_\Lf - y_\Lb$, where $y_\Lf$ is the rapidity of the jet with the greatest rapidity and $y_\Lb$ is the rapidity of the jet with the smallest rapidity. We consider only events in which the average of the transverse momenta of the most forward and backward jets is sufficiently large: $(|\bm p_{\Lf\perp}| + |\bm p_{\Lb\perp}|)/2 > 60 \GeV$. In these events, we measure the total number $N$ of jets. In standard events, the hard scattering is gluon-gluon scattering to produce jets with $|\bm p_{\perp}| > 60 \GeV$. Then $N$ should usually be 2. Initial state radiation can produce a 30 GeV jet in the rapidity range between the jets produced by the hard scattering, but this should be rather rare. In the cut pomeron picture, when $\Delta y$ is large, there is the possibility of producing extra jets in the rapidity range between the jets with extremal rapidities. Thus if we plot the average value $\langle N \rangle$ of the number of jets versus $\Delta y$, we can expect to see a rising curve.

\begin{figure}
\centerline{\hskip 1 cm \includegraphics[width=9.0cm]{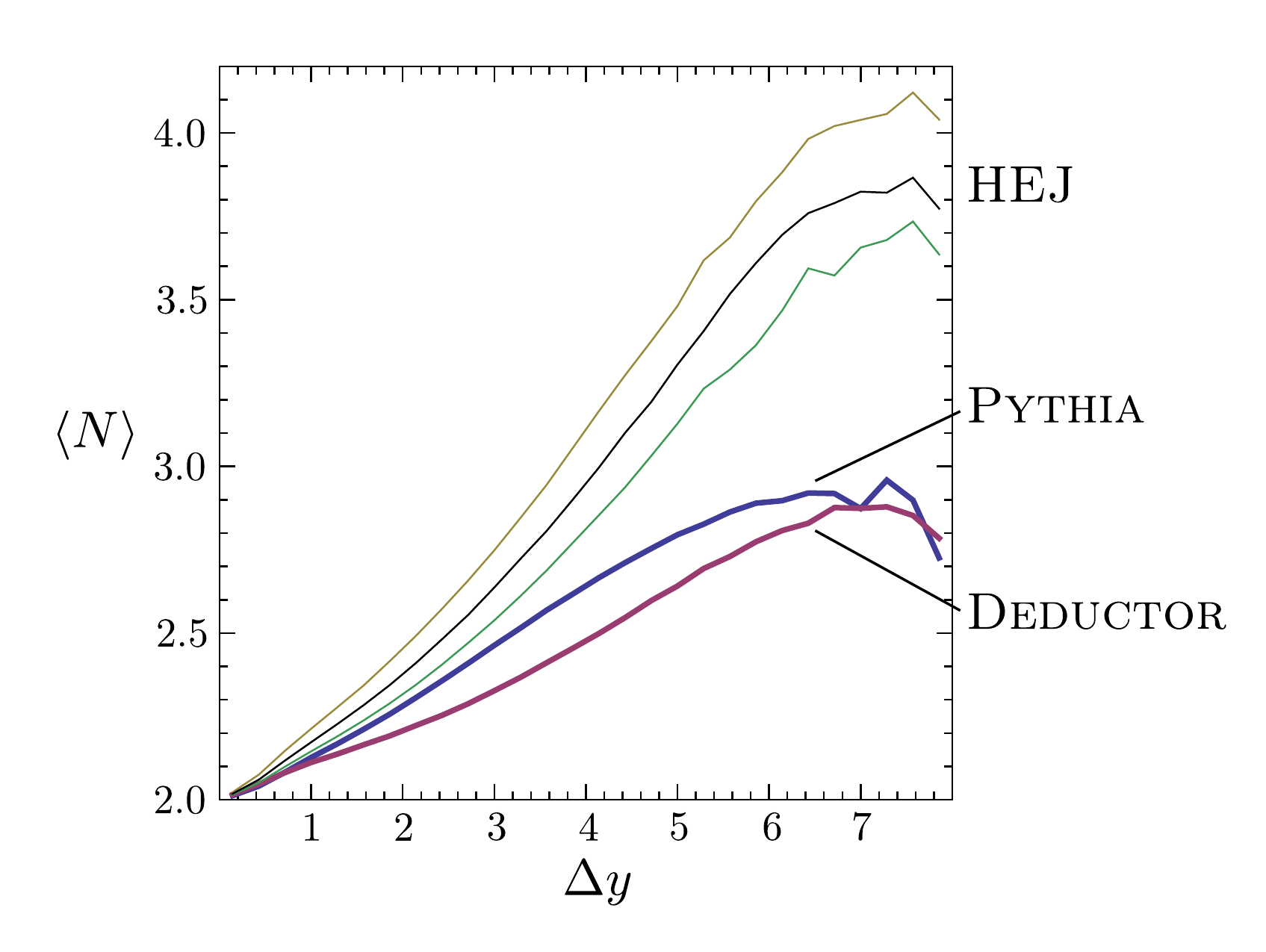}}
\caption{Average number $\langle N \rangle$ of jets versus the rapidity difference $\Delta y$ between the jet with the greatest rapidity and the jet with the least rapidity. All jets must have $|\bm p_{i\perp}| > 30 \GeV$ and the average of the transverse momentum of the two jets with extremal rapidity must be at least $60 \GeV$. The upper three curves are $\langle N \rangle$ according to \textsc{High Energy Jets}, with three choices within this program, as reported in figure~11 of ref.~\cite{HEJ}. The middle curve in the range $5 < \Delta y < 7$ is the result from \textsc{Pythia}. The bottom curve in this range is the result from \textsc{Deductor}.
}
\label{fig:bfkl}
\end{figure}

Plots of $\langle N \rangle$ versus $\Delta y$ are shown in figure \ref{fig:bfkl}. There are five curves. The lowest was generated with \textsc{Deductor}. With \textsc{Deductor}, the production of extra jets between the jets with extremal rapidities is not rare: the average number, $\langle N \rangle - 2$, of extra jets rises with $\Delta y$ to a value of more than 1/2. The next higher curve in this range was generated with \textsc{Pythia} (version 8.176) \cite{Pythia}. The \textsc{Pythia} curve is somewhat higher than the \textsc{Deductor} curve, although, since the \textsc{Pythia} shower is $k_\LT$ ordered, one would think that the \textsc{Pythia} curve should be lower. One effect that tends to raise the \textsc{Pythia} curve is that the value of $\as$ for initial state radiation is large: $\as(M_\LZ^2) = 0.137$ compared to $\as(M_\LZ^2) = 0.118$ in \textsc{Deductor}.\footnote{In \textsc{Deductor}, $\as$ is evaluated at a scale $\lambda_\LR k_\LT^2$, where $\lambda_\LR \approx 0.4$ \cite{deductor}. Thus a more accurate comparison is to $\as(\lambda_\LR M_\LZ^2) = 0.126$ for \textsc{Deductor}.}  The next three curves were generated by \textsc{High Energy Jets} and were taken from figure~11 of ref.~\cite{HEJ}. The three curves correspond to three different choices of parameters within \textsc{High Energy Jets}. We see that the \textsc{High Energy Jets} curves lie substantially above the \textsc{Pythia} and \textsc{Deductor} curves. 

By construction, \textsc{Deductor} allows the transverse momentum of produced jets to decrease and then increase as the initial state shower proceeds over a large rapidity range. The probability that this happens is controlled by Sudakov factors that are constructed to preserve the cross section for the hard event. To represent cut pomeron physics, this may not be what one wants.  We see in figure \ref{fig:bfkl} that although the $\Lambda$ ordering in \textsc{Deductor} allows the production of extra high $k_\LT$ jets, the probability for doing so is not as high as it is in \textsc{High Energy Jets}. We expect that with implementation of more physically motivated Sudakov factors, the \textsc{Deductor} curve may be modified for large $\Delta y$.

\section{Ending the shower}

The shower time that we use, eqs.~(\ref{eq:tdef})  and (\ref{eq:showertime}), is well suited for determining the relative ordering of two splitting vertices. However, at some point in shower evolution, the splittings are too soft for perturbation theory to be reliable. Thus one needs to end the perturbative shower and substitute a non-perturbative model.\footnote{The shower in ref.~\cite{deductor} does not have a non-perturbative hadronization model. We anticipate providing an interface to the string model using \textsc{Pythia}.} The simplest thing to do would be to stop the shower at a fixed value of $\Lambda$. However, that is not sensible because of the factor $1/p_i\!\cdot\! Q_0$ in $\Lambda^2$. This factor is not invariant under boosts in the direction of $p_i$. Consider potential splittings of two partons with the same virtuality $|p_i^2 - m_i^2|$ and the same splitting variable $\z$. With a fixed cutoff on $\Lambda$, a parton that is part of a moderate momentum jet might be allowed to split while a parton that is part of a very high momentum jet is not allowed to split.

Instead, we veto all splittings in which $k_\LT^2 < k_{T,\rm min}^2$ where $k_{T,\rm min}^2$ is, say, 1 GeV.\footnote{Here, to be precise, for a final state splitting, $\bm k_\LT$ is the part of the momentum of one daughter parton relative to the mother parton. For an initial state splitting, $\bm k_\LT$ is part of the momentum of the parton 2 in section~\ref{sec:initialstatesplittings} relative to the direction of initial state parton 1. In each case, we use a reference frame in which $Q_0$ has only a time component.} Neglecting all masses, $k_\LT^2 = z (1-z) |p_i^2|$ for final state splittings, so this cut keeps the virtuality, $z$, and $(1-z)$ from being too small. For initial state splittings with zero masses, $k_\LT^2 = (1-z) |q_i^2|$, so this cut keeps the virtuality and $(1-z)$ from being too small. Since the end of \textsc{Pythia} showers is based on $k_\LT^2$, this choice facilitates matching to the \textsc{Pythia} model of hadronization.

\section{Conclusions}
\label{sec:conclusions}

In a parton shower event generator, one assigns a variable $V_i^2$ to parton splittings for the purpose of ordering the splittings. If the splitting of parton $i$ comes before the splitting of parton $j$ then $V_i^2 > V_j^2$. Often, as in \textsc{Deductor}, $V_i^2$ is related to the hardness of the splitting in the sense that $V_i^2 \to 0$ when the daughter partons become collinear or when one becomes soft.

This hardness ordering criterion allows a rather wide range of choices for $V_i^2$. We propose a more restrictive criterion based on factoring soft interactions from hard interactions. Recall that when one wants to calculate an infrared safe inclusive cross section, one can factor the hardest interaction from parton distribution functions and from softer interactions. In our view, a parton shower event generator works the same way, but on many scales. First the hardest interaction is factored from the rest, then the next hardest interaction, and so forth. We ask that this work on a graph-by-graph basis in a physical gauge (to leading order in the splitting functions). With that demand, we find that $V_i^2$ should be $\Lambda_i^2$, where
\begin{equation}
\begin{split}
\label{eq:showertimeencore}
\Lambda_i^2 ={}& \frac{p_i^2 - m_i^2}{2p_i\cdot Q_0} 
\,Q_0^2
\hskip 1.4 cm {\rm final\ state\ parton}
\;,
\\
\Lambda_i^2 ={}& \frac{|p_i^2 - m_i^2|}{2\eta_i\,p_\LA\cdot Q_0} 
\,Q_0^2
\hskip 1cm {\rm initial\ state\ parton}
\;.
\end{split}
\end{equation}
Here $\eta_i$ is the momentum fraction for an initial state parton from hadron A and $Q_0$ is a fixed timelike vector, which we take to be the sum of the momenta of the final state particles at the hardest interaction. We use this ordering parameter in \textsc{Deductor}.

We investigated in section \ref{sec:kT} the relation of $\Lambda$ ordering to $k_\LT$ ordering, at least for the case of massless final state partons and for just two successive splittings. The ordering of vertices according to $\Lambda$ and $k_\LT$ can be different in the case of wide angle emission of soft gluons. Of course, wide angle, soft gluon emission is important in a gauge theory like QCD. However, wide angle, soft gluon emission has important color coherence properties. Taking color coherence into account, we find that $k_\LT$ ordered emissions, treated with the parton shower approximation of on-shell daughter partons and summed over graphs, gives approximately the same result as $\Lambda$ ordered emissions. Indeed, angle ordered parton showers, by design, also reorder emissions to get an equivalent result to hardness ordered emissions. Thus parton shower algorithms are surprisingly robust against changes of the ordering prescription, at least for final state splittings to the level that we have investigated the question. Nevertheless, we prefer the ordering parameter of eq.~(\ref{eq:showertimeencore}) because, with this choice, factorization works graph by graph.

We investigated in section \ref{sec:InitialState} the relation of $\Lambda$ ordering to $k_\LT$ ordering for the initial state shower in hadron-hadron collisions. We found that $\Lambda$ ordering makes available a wider phase space in the case of a series of splittings with small values of the momentum fraction variable $z$. The wider phase space incorporates the phase space associated with cut pomeron exchange. We have not investigated the Sudakov factors that should be associated with such emissions. Currently in \textsc{Deductor} we use only the standard, probability conserving, Sudakov factors. Then we find that the additional phase space is not often filled.
\acknowledgments{
This work was supported in part by the United States Department of Energy and by the Helmoltz Alliance ``Physics at the Terascale." We thank Jennifer Smillie and Jeppe Andersen for sending us results from \textsc{High Energy Jets}. We thank Hannes Jung for helpful discussions about small $x$ effects.
}

\appendix
\section{Momentum conservation for initial state splitting}
\label{sec:Appendix}

We saw in section~\ref{sec:momentumconservation} how momentum conservation is maintained in generating the initial state shower when all partons are massless. In this appendix, we record the needed Lorentz transformation when the partons are allowed to have non-zero masses.

Let the momenta of the incoming hadrons, treated as massless, be $p_\LA$ and $p_\LB$, defined so that $s = 2 p_\LA\cdot p_\LB$. Let the momentum fractions of the incoming partons be $\eta_\La$ and $\eta_\Lb$. The incoming partons, with flavors $a$ and $b$, may have non-zero masses, so we define their momenta to be
\begin{equation}
\begin{split}
p_\La ={}& \eta_\La p_\LA + \frac{m(a)^2}{\eta_\La \eta_\Lb s}\,\eta_\Lb p_\LB
\;,
\\
p_\Lb ={}& \eta_\La p_\LB + \frac{m(b)^2}{\eta_\La \eta_\Lb s}\,\eta_\La p_\LA
\;.
\end{split}
\end{equation}
We can simplify our notation by defining massless vectors
\begin{equation}
\begin{split}
n_\La ={}& \eta_\La p_\LA 
\;,
\\
n_\Lb ={}& \eta_\La p_\LB 
\;,
\end{split}
\end{equation}
and dimensionless mass squared variables
\begin{equation}
\label{eq:nufdef}
\nu(f) = 
\frac{m(f)^2}{\eta_\La \eta_\Lb s}
\;.
\end{equation}
Then
\begin{equation}
\begin{split}
p_\La ={}& n_\La + \nu(a)\,n_\Lb
\;,
\\
p_\Lb ={}& n_\Lb + \nu(b)\,n_\La
\;.
\end{split}
\end{equation}
We can use the vectors $n_\La$ and $n_\Lb$ as two basis vectors. The projection operator onto the space transverse to $n_\La$ and $n_\Lb$ is given by
\begin{equation}
g_\perp^{\mu\nu} = 
 g^{\mu\nu}
 - \frac{\displaystyle n_{\La}^{\mu}n_{\Lb}^{\nu} 
 + n_{\Lb}^{\mu} n_{\La}^{\nu}}
{\displaystyle n_{\La}\!\cdot\!n_{\Lb}}
\;.
\end{equation}

Now suppose that parton ``a'' splits, in the sense of backward evolution. The new incoming parton has flavor $\hat a$, mass $m(\hat a)$, and momentum fraction $\hat \eta_\La$. We define the momentum fraction of the splitting as $z = \eta_\La/\hat\eta_\La$.\footnote{In generating the shower, we define the splitting variable $z = \eta_\La/\hat\eta_\La$. Elsewhere in this paper, we use $z$ to denote a momentum fraction ratio in several contexts. The precise value of $z$ depends on which partons are approximated as being on shell and which are allowed to be off shell. We hope that this does not cause confusion.} The momentum of the new initial state parton is then 
\begin{equation}
\hat p_\La = \frac{1}{z}\,n_\La + z\nu(\hat a)\,n_\Lb
\;.
\end{equation}
A new final state particle, labelled $m+1$, is emitted. Its momentum has the form
\begin{equation}
\label{eq:xaxbdef}
\hat p_{m+1} = 
x_\La n_\La
+ x_\Lb n_\Lb
+ k_\perp
\;,
\end{equation}
where the magnitude of the transverse momentum is fixed by
\begin{equation}
|k_\perp^2| = 2 x_\La x_\Lb\,  n_\La\cdot n_\Lb - m(f_{m+1})^2
\;.
\end{equation}
The azimuthal angle, $\phi$, of $k_\perp$ is another of the splitting variables besides $z$. The third splitting variable is the dimensionless virtuality
\begin{equation}
y = -\frac{(\hat p_{\La} - \hat p_{m+1})^2 - m^2(a)}{2 n_\La\cdot n_\Lb}
\;.
\end{equation}
Typically, $y \ll 1$.

In order to conserve momentum, we need to adjust the momenta of all of the previously created final state particles,
\begin{equation}
p_i^\mu \to \hat p_i^\mu = \Lambda^{\mu}_{\ \nu}\,p_i^\nu
\hskip 1 cm i = 1,\dots,m
\;.
\end{equation}
This works provided 
\begin{equation}
\begin{split}
\label{eq:xaxbResult}
x_\La ={}&
\frac{1/z
-1 
- y \left(1+z\,\nu(b)\right)
- \left(1-z\right)\nu(a) \nu(b)  
- z\,\nu(f_{m+1})\nu(b) }
{1 - z^2
\nu(\hat a)\nu(b)}
\;,
\\
x_\Lb ={}&
z\,\frac{y  + \nu(f_{m+1}) - \nu(a)
+(1+y)\,z\,\nu(\hat a)
+ z\,\nu(\hat a)\nu(b)
\left(\nu(a) 
- z\,\nu(\hat a)
\right)}
{1 - z^2\, \nu(\hat a)\nu(b)}
\;.
\end{split}
\end{equation}

The needed Lorentz transformation starts with a small boost along the $z$-axis
\begin{equation}
\label{eq:boostparallel}
\Lambda_{\parallel}^{\mu\nu}(\omega) = g_\perp^{\mu\nu}
+\frac{ 
   e^{\omega}\,n_{\La}^{\mu}n_{\Lb}^{\nu} 
+ e^{-\omega}\, n_{\Lb}^{\mu} n_{\La}^{\nu}
}
{n_{\La}\!\cdot\!n_{\Lb}}
\;,
\end{equation}
with boost angle $\omega$ given by
\begin{equation}
\label{eq:omegadef}
e^\omega = 
[1 + \nu(b)]^{-1}
\left[
\frac{1}{z}
+ \nu(b)
- x_\La
\right]
\;.
\end{equation}
Then we apply a transverse null plane boost $\Lambda_\perp(v_\perp)$,
\begin{equation}
\label{eq:boostperp}
\Lambda_{\perp}^{\mu\nu}(v_{\perp}) = g^{\mu\nu} 
+\sqrt{\frac{2}{n_{\La}\!\cdot\!n_{\Lb}}}
\left[ v_{\perp}^{\mu}n_{\Lb}^{\nu} - n_{\Lb}^{\mu} v_{\perp}^{\nu}\right]
-v_{\perp}^{2}\,
\frac{n_{\Lb}^{\mu}n_{\Lb}^{\nu}}{n_{\La}\!\cdot\!n_{\Lb}}
\;,
\end{equation}
with boost velocity
\begin{equation}
\label{eq:vdef}
v_{\perp} = - \frac{e^{-\omega}}{[1 + \nu(b)]\sqrt{2 n_{\La}\!\cdot\!n_{\Lb}}}\
k_\perp
\;.
\end{equation}
Our normalization convention for $v_\perp$ here is different from that used in section~\ref{sec:momentumconservation}. We also note that $\Lambda(v_\perp)$ leaves $n_\Lb$ unchanged, but it changes $p_\Lb$ by a small amount, proportional to $\nu(b)$. Thus the interpretation is a little different from that described in section~\ref{sec:momentumconservation} for massless partons. However, the difference in interpretation is of no real consequence.

The complete Lorentz transformation is
\begin{equation}
\begin{split}
\label{eq:boostall}
\Lambda^{\mu\nu}(\omega,v_{\perp}) ={}& g_{\perp}^{\mu\nu}
+
\frac{ 
   e^{\omega}\,n_{\La}^{\mu}n_{\Lb}^{\nu} 
+ e^{-\omega}\, n_{\Lb}^{\mu} n_{\La}^{\nu}
}
{  n_{\La}\!\cdot\!n_{\Lb}}
\\ &
+\sqrt{\frac{2}{n_{\La}\!\cdot\!n_{\Lb}}}
\left[ 
e^{\omega} v_{\perp}^{\mu}n_{\Lb}^{\nu} - n_{\Lb}^{\mu} v_{\perp}^{\nu}\right]
-  e^{\omega}  v_{\perp}^{2}\,\frac{n_{\Lb}^{\mu}n_{\Lb}^{\nu}}
{n_{\La}\!\cdot\!n_{\Lb}}
\;\;.
\end{split}
\end{equation}

The analysis of this appendix follows from applying the choice $y,z,\phi$ of splitting variables to the analysis of section 4.4 of ref.~\cite{NSI} except that we choose the Lorentz transformation  $\Lambda^{\mu}_{\ \nu}$ differently. As noted at eq.~(3.9) of ref.~\cite{NSDrellYan}, the choice that we make here is better adapted to summing logs to evaluate the transverse momentum of Z-bosons produced in the Drell-Yan process.


\end{document}